\documentclass{article} 

\usepackage{graphicx}

\usepackage[utf8]{inputenc}
\newtheorem{theorem}{Theorem}[section]

\usepackage{amsmath,amssymb,amsfonts}
\usepackage{wasysym}

\newcommand{\mq}{m_{bh}}
\newcommand{\Rt}{\mathbb{R}^3}
\newcommand{\Su}{\Sigma}
\newcommand{\dom}{U}

\newcommand{\Si}{\mathcal{R}} 

\newcommand{\ra}{\mathcal{R}_{SY}}

\begin{document}

\title{Geometric inequalities for  black holes}


\author{Sergio Dain\\
Facultad de Matem\'atica, Astronom\'{\i}a y
  F\'{i}sica, FaMAF, \\
Universidad  Nacional de C\'ordoba,\\
  Instituto de F\'{\i}sica Enrique Gaviola, IFEG, CONICET,\\
  Ciudad Universitaria (5000) C\'ordoba, Argentina.
}

\maketitle

\begin{abstract}
  It is well known that the three parameters that characterize the Kerr black
  hole (mass, angular momentum and horizon area) satisfy several important
  inequalities. Remarkably, some of these inequalities remain valid also for
  dynamical black holes. This kind of inequalities play an important role in
  the characterization of the gravitational collapse. They are closed related
  with the cosmic censorship conjecture. In this article recent results in this
  subject are reviewed.

\end{abstract}

\section{Geometric inequalities in General Relativity}
\label{sec:geom-ineq}

\begin{figure}
\begin{center}
\includegraphics[width=3cm]{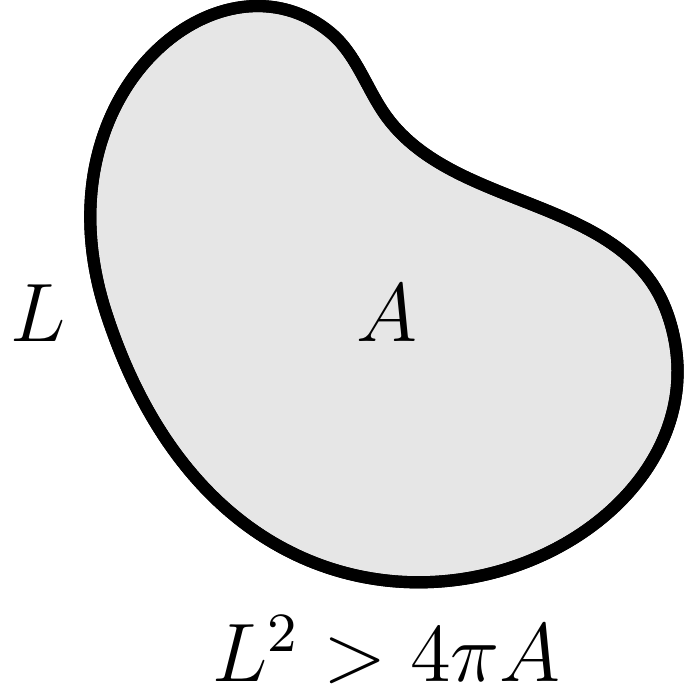}
\hspace{1.5cm}
\includegraphics[width=3cm]{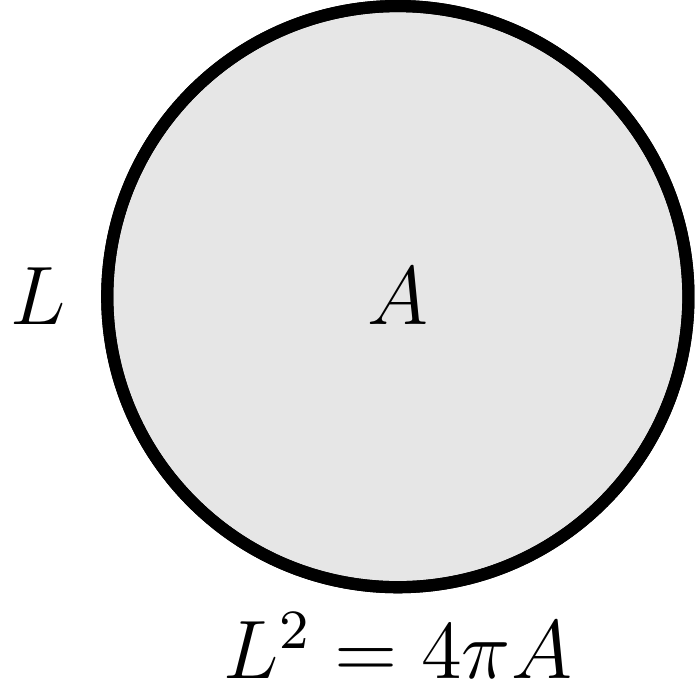}
\end{center}
\label{fig:1}
\caption{The isoperimetric inequality. On the left an arbitrary curve, where the strict inequality holds. On the right the circle, where the equality holds.}
\end{figure}

A classical example of a geometric inequality is the isoperimetric inequality for closed plane curves given by
\begin{equation}
  \label{eq:54}
  L^2 \geq 4\pi A\quad (=\text{ circle}),
\end{equation}
where $A$ is the area enclosed by a curve $C$ of length $L$. In (\ref{eq:54})
equality holds if and only if $C$ is a circle, see figure \ref{fig:1}. For a
review on this subject see \cite{Osserman78}.
The inequality (\ref{eq:54}) applies to complicated geometric objects
(i.e. arbitrary closed planar curves).  The equality in (\ref{eq:54}) is achieved
only for an object of ``optimal shape'' (i.e. the circle) which is described by
few parameters (in this case only one: the radius). Moreover, this object has a
variational characterization: the circle is uniquely characterized by the
property that among all simple closed plane curves of given length $L$, the
circle of circumference $L$ encloses the maximum area.

General Relativity is a geometric theory, hence it is not surprising that
geometric inequalities appear naturally in it.  Many of these inequalities are
similar in spirit as the isoperimetric inequality (\ref{eq:54}). In particular,
all the geometric inequalities discussed in this article will have the same
structure as (\ref{eq:54}): the inequality applies for a rich class of objects
and the equality only applies for an object of ``optimal shape'' (always
indicated in parenthesis as in (\ref{eq:54})). This object, like the circle,
can be described by few parameters and it has also a variational
characterization.

However, General Relativity is also a physical theory.  It is often the case
that the quantities involved have a clear physical interpretation and the
expected behavior of the gravitational and matter fields often suggests
geometric inequalities which can be highly non-trivial from the mathematical
point of view.  The interplay between physics and geometry gives to geometric
inequalities in General Relativity their distinguished character. These
inequalities relate quantities that have both a physical interpretation and a
geometrical definition.

The plan of this article follows this interplay between physics and
mathematics.  In section \ref{sec:physical-picture} we present the physical
motivations for the black holes geometric inequalities. In section
\ref{sec:theorems} we summarize some theorems where these inequalities have
been recently proved. Finally, in section \ref{sec:open-problems-recent} we
list relevant open problems and we also describe recent results on geometric
inequalities for bodies.

\section{Physical picture}
\label{sec:physical-picture}
An important example of a geometric inequality is the positive mass theorem. 
Let $m$ be the total ADM mass on an asymptotically flat complete initial data
such that the dominant energy condition is satisfied. Then we have
\begin{equation}
  \label{eq:1}
0\leq m\quad  (=\text{ Minkowski}).  
\end{equation}
The mass $m$ is a pure geometrical quantity
\cite{Arnowitt62}\cite{Bartnik86}\cite{chrusciel86}.  However, from the
geometrical mass definition, without the physical picture, it would be very
hard even to conjecture the inequality (\ref{eq:1}). In fact the proof of the
positive mass theorem turns out to be very subtle
\cite{Schoen79b}\cite{Schoen81}\cite{witten81}.

A key assumption in the positive mass theorem is that the matter fields should
satisfy an energy condition.  This condition is expected to hold for all
physically realistic matter. This kind of general properties which do not
depend very much on the details of the model are not easy to find for a
macroscopic object.  And hence it is difficult to obtain simple and general
geometric inequalities among the parameters that characterize ordinary
macroscopic objects. Black holes represent a unique class of very simple
macroscopic objects and hence they are natural candidates for geometrical
inequalities. Nevertheless, in section \ref{sec:open-problems-recent} we will
present also a  geometric inequality valid for ordinary bodies.

The black hole uniqueness theorem ensures that stationary black holes in vacuum
are characterized by the Kerr exact solution of Einstein equations \footnote{It
  is worth mention that important aspects of the black hole uniqueness problem
  remain still open, see recent review article \cite{lrr-2012-7} and reference
  therein.}. For simplicity we will not consider the electromagnetic field in
this article, however most of the results presented here can be generalized to
include that case.

\begin{figure}
\begin{center}
\includegraphics[width=0.7\textwidth]{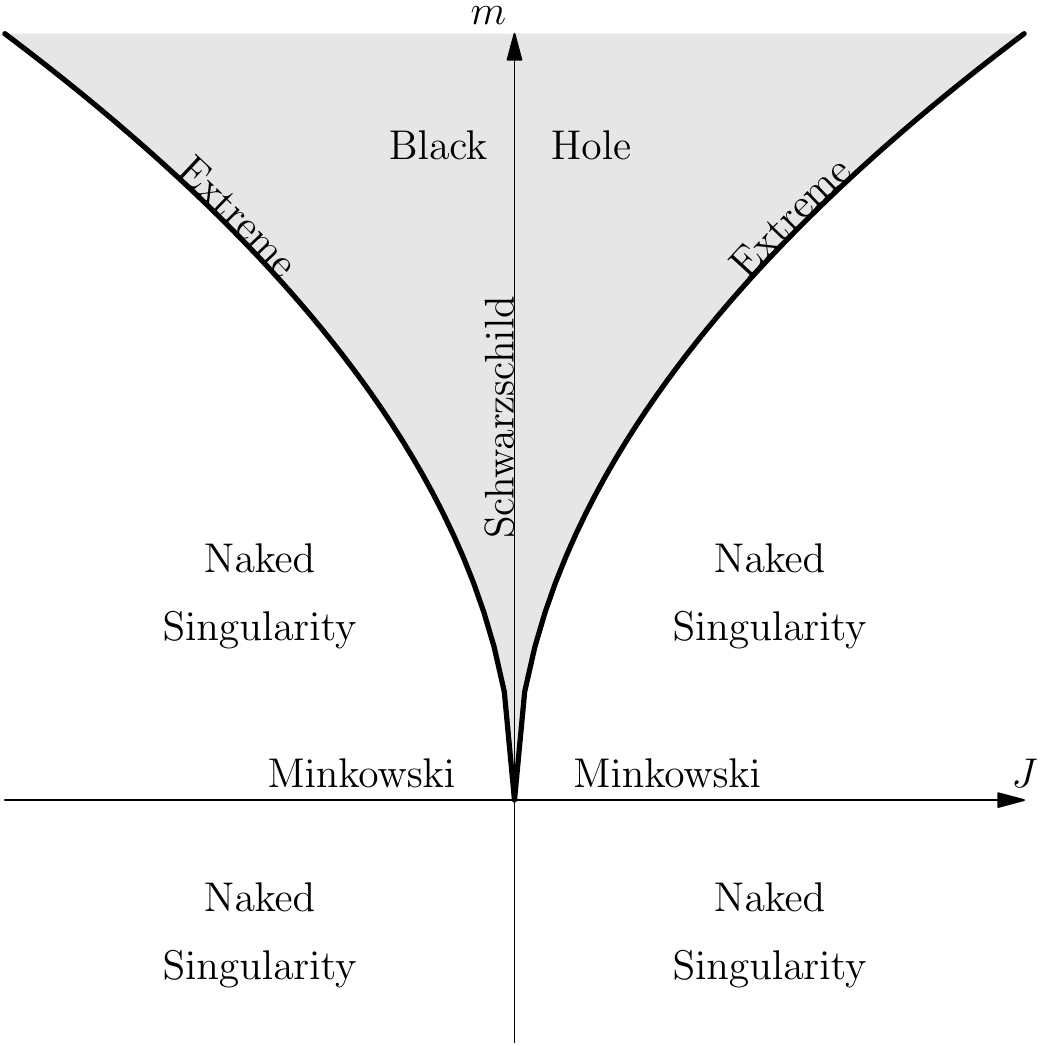}
\end{center}
\caption{A point in this graph is a Kerr solution with parameters $m$ and
  $J$. The horizontal axis where $m=0$ is Minkowski space.  The Schwarzschild
  solution is given by the vertical axis where $J=0$.  In the gray region the
  parameters satisfy the inequality (\ref{eq:2}) and hence the Kerr solution
  describe a black hole. The boundary of this region is given by the equality
  in (\ref{eq:2}), these solutions are called extreme black holes. In the white
  region, excluding the horizon axis, the Kerr solution contains a naked
  singularity. That includes also the negative mass region. }
\label{fig:2}
\end{figure}

It is somehow remarkable that the same family of solutions of Einstein
equations that describe the unique stationary black hole (i.e. the Kerr metric)
also describe naked singularities. In effect, the Kerr metric depends on two
parameters: the mass $m$ and the angular momentum $J$.  This metric is a
solution of Einstein vacuum equations for any choice of the parameters $m$ and
$J$. However, it represents a black hole if and only if the following
remarkably inequality holds
\begin{equation}
    \label{eq:2}
\sqrt{|J|}\leq  m.    
\end{equation}
Otherwise the spacetime contains a naked singularity.  Figure \ref{fig:2} shows
the parameter space of the Kerr solution. Extreme black holes are defined by
the equality in (\ref{eq:2}). These black holes lie at the boundary between
naked singularities and black holes. For most of the inequalities discussed in
this article, extreme black holes play the role of the circle in the
isoperimetric inequality (\ref{eq:54}): they reach the equality and they
represent objects of ``optimal shape''.

The area of the horizon of the Kerr black hole is given by the simple but very important formula
\begin{equation}
  \label{eq:3}
  A=8\pi \left(m^2+ \sqrt{m^4-J^2}  \right).
\end{equation}
From equation (\ref{eq:3}) we deduce that the  following three geometric inequalities  hold for a Kerr black hole 
\begin{align}
 \sqrt{\frac{A}{16\pi}} &\leq m &(=\text{Schwarzschild}),\label{eq:pen}\\
   \sqrt{|J|} &\leq  m  &(= \text{Extreme Kerr}),\label{eq:mj}\\
    8\pi |J| &\leq  A  &(= \text{Extreme Kerr}).\label{eq:JA}
\end{align} 
As expected from the discussion above, the inequality (\ref{eq:mj}) is needed
to define the black hole horizon area in (\ref{eq:3}): if (\ref{eq:mj}) does
not hold, then the expression (\ref{eq:3}) is not a real number.  We have
listed this inequality again here to emphasize its connection with the other
two in the following discussion.  Inequalities (\ref{eq:pen}) and (\ref{eq:JA})
follow from (\ref{eq:mj}) and (\ref{eq:3}).  Note that these inequalities
relate the three relevant parameters of the Kerr black hole $(m,J,A)$.

Let us discuss the physical meaning of the inequalities (\ref{eq:pen}),
(\ref{eq:mj}) and (\ref{eq:JA}). In the inequality (\ref{eq:pen}), the
difference
\begin{equation}
  \label{eq:4b}
 m-\sqrt{\frac{A}{16\pi}},
\end{equation}
represents the rotational energy of the Kerr black hole. This is the maximum
amount of energy that can be extracted from the black hole by the Penrose
process \cite{Christodoulou70}. When the difference (\ref{eq:4b}) is zero, the
black hole has no angular momentum and hence it is the Schwarzschild black
hole.

From Newtonian considerations, we can interpret the inequality (\ref{eq:mj}) as
follows \cite{Wald71}.  In a collapse the gravitational attraction ($\approx
m^2/r^2$) at the horizon ($r \approx m $) dominates over the centrifugal
repulsive forces ($\approx J^2/mr^3$).

Finally, concerning the inequality (\ref{eq:JA}), the black hole temperature is
given by the following formula
\begin{equation}
  \label{eq:5c}
  \kappa= \frac{1}{4 m} \left(1-\frac{(8\pi J)^2 }{A^2}  \right).
\end{equation}
The temperature is positive if and only if the inequality (\ref{eq:JA})
holds. Moreover the temperature is zero if and only if the equality in
(\ref{eq:JA}) holds and hence the black hole is extreme. 

There exists another relevant geometrical inequality which can be deduced from
the formula (\ref{eq:3})
\begin{equation}
  \label{eq:10}
8\pi \left( m^2-\sqrt{m^4-J^2} \right) \leq  A \quad  (= \text{Extreme Kerr}).
\end{equation}
Remarkably, as it was pointed out in \cite{Khuri:2013wha} for the case of the
electric charge and in \cite{Dain:2013qia} for the present case of angular
momentum, the inequality (\ref{eq:10}) can be deduced purely from the
inequalities (\ref{eq:mj}) and (\ref{eq:JA}) (i.e. without using the equality
(\ref{eq:3})) by simple algebra. Namely
\begin{align}
  \label{eq:11}
  m^2 &= \sqrt{m^4-J^2+J^2},\\
& \leq |J| +\sqrt{m^4-J^2}, \label{eq:11b} \\
& \leq \frac{A}{8\pi}+ \sqrt{m^4-J^2},\label{eq:11c}
\end{align}
where in the line (\ref{eq:11b}) we have used  (\ref{eq:mj}) and in line
(\ref{eq:11c}) we have used (\ref{eq:JA}). In that sense, the inequalities
(\ref{eq:pen}), (\ref{eq:mj}) and (\ref{eq:JA})   are
more fundamental than (\ref{eq:10}). However, the inequality (\ref{eq:10}) is
important by itself since it related with the Penrose inequality with angular
momentum,  see \cite{Khuri:2013wha} \cite{Dain:2013qia}.

We have seen that for stationary black holes the inequalities (\ref{eq:pen}),
(\ref{eq:mj}) and (\ref{eq:JA}) are straightforward consequences of the area
formula (\ref{eq:3}).  

\begin{figure}
\begin{center}
\includegraphics[width=6cm]{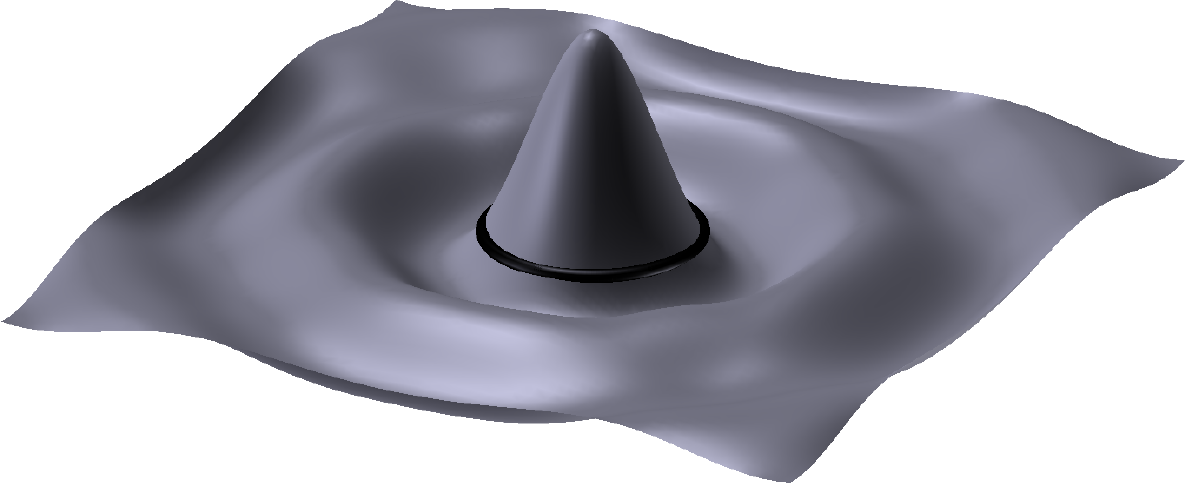} 
\end{center}
\caption{Schematic representation of an initial data for a non-stationary black
  hole. The black ring represents a trapped surface. Outside and inside the
  trapped surface the gravitational field is highly dynamical.}
\label{fig:3}
\end{figure}

However, black holes are in general non stationary, see figure \ref{fig:3}.
Astrophysical phenomena like the formation of a black hole by gravitational
collapse or a binary black hole collision are highly dynamical. For such
systems, the black hole can not be characterized by few parameters as in the
stationary case. In fact, even stationary but non-vacuum black holes have a
complicated structure (for example black holes surrounded by a rotating ring of
matter, see the numerical studies in \cite{Ansorg05}). Remarkably, inequalities
(\ref{eq:pen}), (\ref{eq:mj}) and (\ref{eq:JA}) extend (under appropriate
assumptions) to the fully dynamical regime. Moreover, these inequalities are
deeply connected with properties of the global evolution of Einstein equations,
in particular with the cosmic censorship conjecture.

To discuss the physical arguments that support these inequalities in the
dynamical regime it is convenient to start with the inequality \eqref{eq:JA}.
For a dynamical black hole, the physical quantities that are well defined are
the total ADM mass $m$ of the spacetime and the area $A$ of the black hole
horizon.  The total mass $m$ of the spacetime measures the sum of the black
hole mass and the mass of the gravitational waves surrounding it. In the
stationary case, the mass of the black hole is equal to the total mass of the
spacetime, but this is no longer true for a dynamical black hole.  The mass $m$
is a global quantity, it carries information on the whole spacetime. In
contrast, the area of the horizon $A$ is a quasi-local quantity, it carries
information on a bounded region of the spacetime.

It is well known that the energy of the gravitational field cannot be
represented by a local quantity (i.e. a scalar field). The best one can hope is
to obtain a quasi-local expression. The same applies to the angular
momentum.  In general, it is difficult to find physically relevant quasi-local
quantities like mass and angular momentum (see the review article
\cite{Szabados04}).  However, in axial symmetry, there is a well defined notion
of quasi-local angular momentum: the Komar integral of the axial
Killing vector. Moreover, the angular momentum is conserved in vacuum. That is,
axially symmetric gravitational waves do not carry angular momentum.

Then, for axially symmetric dynamical black holes we have two well defined
quasi-local quantities: the area of the horizon $A$ and the angular momentum
$J$. Note that the inequality \eqref{eq:JA} relates only quasi-local
quantities. 

Using $A$ and $J$ we can define the quasi-local mass for a dynamical black hole
by the Kerr formula (\ref{eq:3}), that is
\begin{equation}
\label{eq:masa}
  \mq= \sqrt{\frac{A}{16\pi}+\frac{4\pi J^2}{A}}.
\end{equation}
This is, in principle, just a definition. Since $\mq$ is given by the Kerr
formula (\ref{eq:3}) it automatically satisfies the inequalities (\ref{eq:mj})
and (\ref{eq:pen}).  However, the relevant question is: does $\mq$ describes
the quasi-local mass of a non-stationary black hole? This question is closed
related to the validity of the inequality (\ref{eq:JA}) in the dynamical
regime. In order to answer it let us analyze the evolution of $\mq$.

\begin{figure}
\begin{center}
\includegraphics[width=6cm]{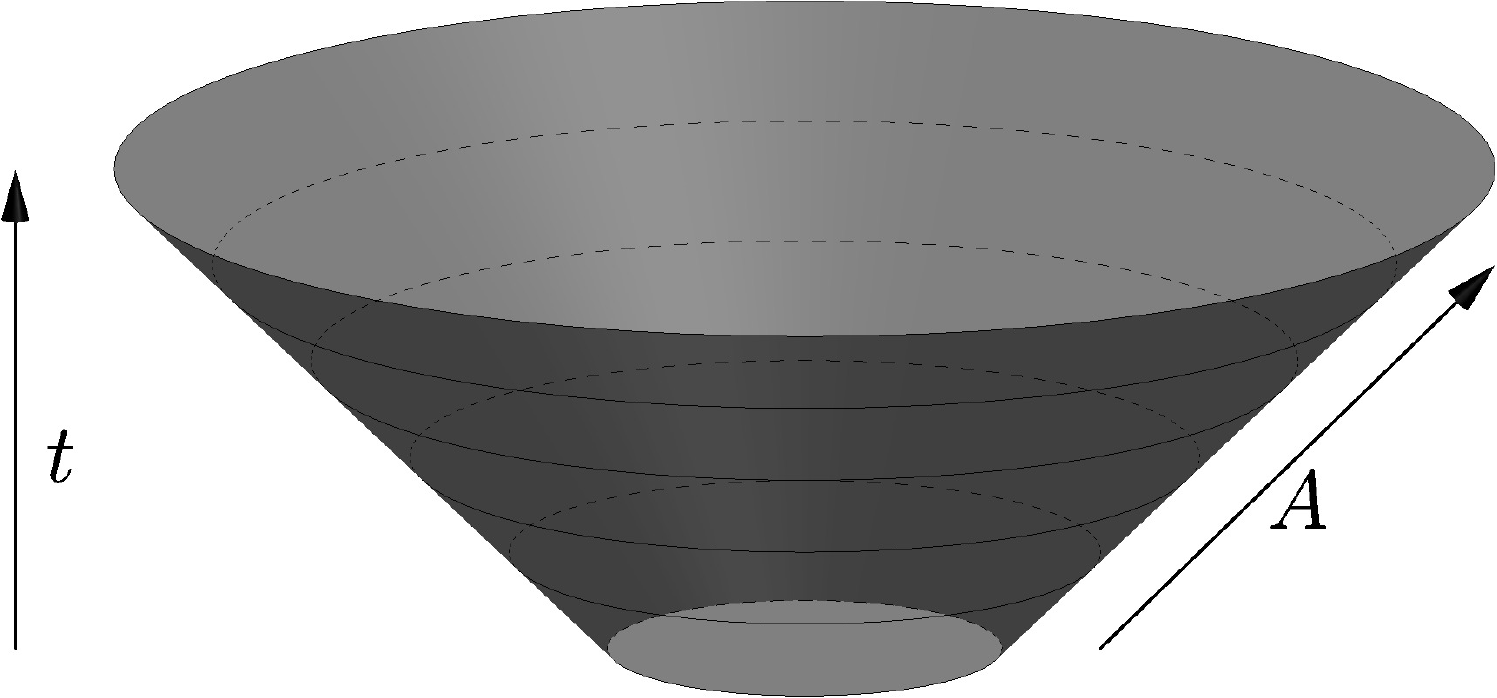}
\end{center}
\caption{The area theorem. The horizon area of a dynamical black hole increase
  with time.}
\label{fig:4}
\end{figure}

For a dynamical black hole, by the area theorem, we know that the horizon area
$A$ increase with time, see figure \ref{fig:4}.
In general, the quasi-local mass of the black hole is not expected to be a
monotonically increasing quantity. Energy can be extracted from a rotating black
hole by the Penrose process. However, if we assume axial symmetry then the
angular momentum will be conserved at the quasi-local level. On physical
grounds, one would expect that in this situation the quasi-local mass of the
black hole should increase with the area, since there is no mechanism at the
classical level to extract mass from the black hole. In effect, the Penrose
process involves an interchange of angular momentum between the black hole and the
exterior. But the angular momentum transfer is forbidden in axial
symmetry. Then,  both the area $A$ and the quasi- local mass
$\mq$ should monotonically increase with time in axial symmetry. 

Let us take a time derivative of $\mq$. To analyze this, it is illustrative to
write down the complete differential, namely the first law of thermodynamics
\begin{equation}
\label{eq:mq}
  \delta \mq=  \frac{\kappa}{8 \pi} \delta A + \Omega_H \delta J,
\end{equation}
where
\begin{equation}
  \label{eq:7}
  \kappa= \frac{1}{4\mq} \left(1-\frac{(8\pi J)^2 }{A^2} \right),\quad
  \Omega_H=\frac{4\pi J}{A \,\mq}.
\end{equation}
In equation (\ref{eq:mq}) we have followed the standard notation for the
formulation of the first law; we emphasize, however, that in our context this
equation is a trivial consequence of \eqref{eq:masa}. In axial symmetry $\delta
J=0$ and hence we obtain
\begin{equation}
  \delta  \mq=  \frac{\kappa}{8 \pi} \delta A.
\end{equation}
By the area theorem we have
\begin{equation}
  \delta A \geq 0.
\end{equation}
Then $\delta \mq\geq 0$ if and only if $\kappa \geq 0$, that is $\delta \mq\geq
0$ if and only if the inequality (\ref{eq:JA}) holds. Then, it is natural to
conjecture that this inequality should be satisfied for any axially symmetric
black hole. If the horizon violates (\ref{eq:JA}), then in the evolution the
area will increase but the mass $\mq$ will decrease. This will indicate that
the quantity $\mq$ does not have the desired physical meaning. Also, a rigidity
statement is expected. Namely, the equality in (\ref{eq:JA}) is reached only by
the extreme Kerr black hole where $\kappa=0$.

This inequality provides a remarkable quasi-local measure of how far a
dynamical black hole is from the extreme case, namely an `extremality criteria'
in the spirit of \cite{Booth:2007wu}, although restricted only to axial
symmetry.  In the article \cite{Dain:2007pk} it has been conjectured that,
within axially symmetry, to prove the stability of a nearly extreme black hole
is perhaps simpler than a Schwarzschild black hole. It is possible that this
quasi-local extremality criteria will have relevant applications in this
context.  Note also that the inequality \eqref{eq:JA} allows to define, at
least formally, the positive temperature of a dynamical black hole $\kappa$ by
the formula (\ref{eq:7}) (see Refs. \cite{Ashtekar03} \cite{Ashtekar02} for a
related discussion of the first law in dynamical horizons).  If inequality
\eqref{eq:JA} holds, then $\mq$ defines a non-trivial quantity that increase monotonically with time, like the  black hole area $A$.

It is important to emphasize that the physical arguments presented above in
support of \eqref{eq:JA} are certainly weaker in comparison with the ones
behind the Penrose inequalities that support the inequalities (\ref{eq:pen})
and (\ref{eq:mj}) that we will discuss bellow.  A counter example of
any of these inequality will prove that the standard picture of the
gravitational collapse is wrong. On the other hand, a counter example of
\eqref{eq:JA} will just prove that the quasi-local mass \eqref{eq:mq} is not
appropriate to describe the evolution of a non-stationary black hole.  One can
imagine other expressions for quasi-local mass, may be more involved, in axial
symmetry.  On the contrary, reversing the argument, a proof of \eqref{eq:JA}
will certainly suggest that the mass \eqref{eq:mq} has physical meaning for
non-stationary black holes as a natural quasi-local mass (at least in axial
symmetry). Also, the inequality \eqref{eq:JA} provide a non trivial control of
the size of a black hole valid at any time.

In a seminal article Penrose \cite{Penrose73} proposed a remarkably physical
argument that connects global properties of the gravitational collapse with
geometric inequalities on the initial conditions. That argument lead to the
well known Penrose inequality (\ref{eq:pen}) for dynamical black holes (without
any symmetry assumption). In the following we review this argument imposing
axial symmetry, where angular momentum is conserved. And, more important, we
include a relevant new ingredient: we assume that the inequality (\ref{eq:JA})
holds.

We will assume that the following statements hold in a gravitational collapse:
\begin{itemize}

\item[(i)]  Gravitational collapse results in a black hole (weak cosmic
censorship). 

\item[(ii)]  The spacetime settles down to a stationary final
state. We will further assume that at some finite time all the matter have
fallen into the black hole and hence the exterior region is vacuum. 
\end{itemize}
Conjectures (i) and (ii) constitute the standard picture of the gravitational
collapse. Relevant examples where this picture is confirmed (and where the role
of angular momentum is analyzed) are the collapse of neutron stars studied
numerically in \cite{Baiotti:2004wn} \cite{Giacomazzo:2011cv}.
 
\begin{figure}
\begin{center}
\includegraphics[width=0.5\textwidth]{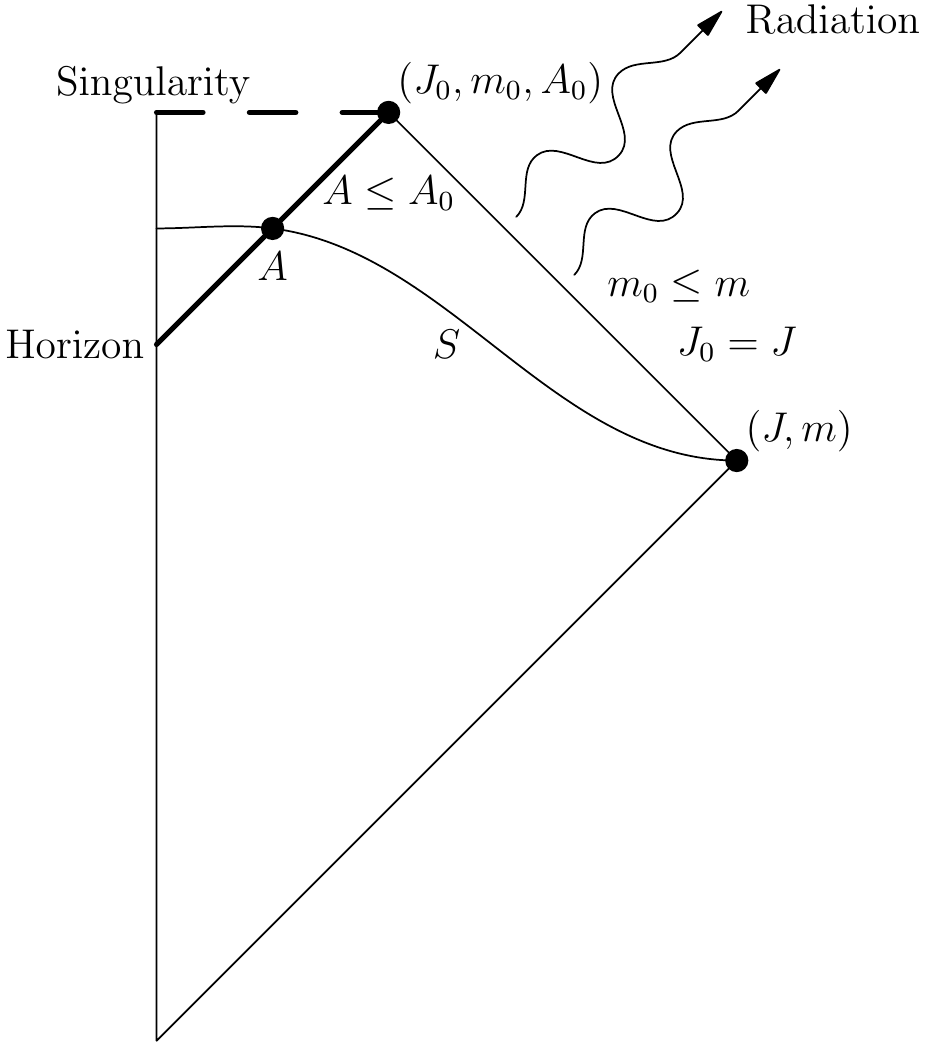}
\end{center}
\caption{The Penrose diagram of a gravitational collapse. The initial Cauchy
  surface is denoted by $S$. The area $A$ increase along the event horizon. The
  mass $m$ decrease along null infinity. We have assumed axial symmetry and
  hence the angular momentum remains constant along null infinity $J=J_0$.}
\label{fig:5}
\end{figure}

The black hole uniqueness theorem implies that the final stationary state
postulated in (ii) is given by the Kerr black hole.  Let us denote by $m_0,
J_0, A_0$, respectively, the mass, angular momentum and horizon area of the
remainder Kerr black hole.  Penrose argument runs as follows. Take a Cauchy
surface $S$ in the spacetime such that the collapse has already occurred. This
is shown in figure \ref{fig:5}.
Let $\Su$ denotes the intersection of the event horizon with the Cauchy surface
$S$ and let $A$ be its area.  Let $(m, J)$ be the total mass and angular
momentum at spacelike infinity. These quantities can be computed from the
initial surface $S$. By the black hole area theorem we have that the area of
the black hole increase with time and hence
\begin{equation}
  \label{eq:15}
  A_0\geq A.
\end{equation}
Since gravitational waves carry positive energy, the total mass of the
spacetime should be bigger than the final mass of the remainder Kerr black hole
\begin{equation}
  \label{eq:4}
  m\geq m_0.
\end{equation}
The difference $m-m_0$ is the total amount of gravitational radiation emitted
by the system. 

To related the initial angular momentum $J$ with the final angular momentum
$J_0$ is much more complicated. Angular momentum is in general
non-conserved. There exists no simple relation between the total angular
momentum $J$ of the initial conditions and the angular momentum $J_0$ of the
final black hole. For example, a system can have $J=0$ initially, but collapse
to a black hole with final angular momentum $J_0\neq 0$. We can imagine that on
the initial conditions there are two parts with opposite angular momentum, one
of them falls in to the black hole and the other escape to infinity. Axially
symmetric vacuum spacetimes constitute a remarkable exception because the
angular momentum is conserved. In that case we have
\begin{equation}
  \label{eq:59}
  J=J_0.
\end{equation}
For a discussion of this conservation law in detail see \cite{dain12} and
reference therein.

We have assumed that the inequality \eqref{eq:JA} holds, then by the discussion
above we have that the quasi-local mass $\mq$ increase with time, that is
\begin{equation}
  \label{eq:8b}
  \mq \leq m_0.
\end{equation}
We emphasize that this inequality is highly non-trivial. The quantity $\mq$ is
computed on the initial surface $S$, in contrast to compute $m_0$ we need to
known the whole spacetime. Using (\ref{eq:8b}) and (\ref{eq:4}) we finally
obtain
\begin{equation}
  \label{eq:6}
\sqrt{\frac{A}{16\pi}+\frac{4\pi J^2}{A}}=  \mq \leq m. 
\end{equation}
This inequality has the natural interpretation that the mass of the black hole
$\mq$ should always be smaller than the total mass of the spacetime $m$. The
inequality (\ref{eq:6}) represents a generalization of the Penrose inequality
with angular momentum. This inequality implies
\begin{equation}
\label{eq:mjd}
  \sqrt{|J|}\leq m.
\end{equation}
In fact, the inequality \eqref{eq:mjd} can be deduced directly by the same
heuristic argument without using the area theorem. It depends only on the
following assumptions
\begin{itemize}
\item Gravitational waves carry positive energy.

\item Angular momentum is conserved in axial symmetry.

\item In a gravitational collapse the spacetime settles down to a final Kerr
  black hole. 
\end{itemize}

Let us summarize the discussion of this section.  For an axially symmetric,
dynamical black hole, the following two geometrical inequalities are expected
  \begin{align}
    8\pi |J| & \leq A \quad (=\text{Extreme Kerr horizon}), \label{eq:JAd}\\
    \sqrt{\frac{A}{16\pi}+\frac{4\pi J^2}{A}} & \leq m \quad (=\text{Kerr black
      hole}). \label{eq:pendj}
  \end{align}
  The inequality \eqref{eq:JAd} is quasi-local and the inequality
  \eqref{eq:pendj} is global. The global inequality \eqref{eq:pendj} implies
  the following two inequalities
\begin{align}
  \sqrt{\frac{A}{16\pi}} &\leq m \quad (=\text{Schwarzschild}),\label{eq:penddd}\\
\sqrt{|J|} &\leq m. \quad (=\text{extreme Kerr black hole}).\label{eq:mjdd}
\end{align}
 That is: 
\begin{quote}
  \emph{The three geometrical inequalities (\ref{eq:pen}), (\ref{eq:mj}) and
  (\ref{eq:JA}) valid for the Kerr black holes are expected to hold also for
  axially symmetric, dynamical black holes.}
\end{quote}

The Penrose inequality \eqref{eq:penddd} is valid also without the axial
symmetry assumption.  It is important to emphasize that all the quantities
involved in the geometrical inequalities above can be calculated on the initial
surface. For simplicity, we have avoided the distinction between event horizon
and apparent horizons (defined in terms of trapped surfaces) to calculate the
area $A$. This point is important for the Penrose inequality (see the
discussion in \cite{Mars:2009cj}) but not for the other inequalities which are
the main subject of this review.  In particular the horizon area $A$ in
(\ref{eq:JAd}) is the area of an appropriated defined trapped surface.
 
A counter example of the global inequality (\ref{eq:pendj})  
will imply that cosmic censorship  is not true. Conversely a proof of
it  gives indirect evidence of the validity of censorship,
since it is very hard to understand why this highly nontrivial
inequality should hold unless censorship can be thought of as providing
the underlying physical reason behind it. 
 
The inequalities (\ref{eq:pen}), (\ref{eq:mj}) and (\ref{eq:JA}) can be divided
into two groups:

\begin{enumerate}
\item $\sqrt{\frac{A}{16\pi}}\leq m$: the area appears as lower bound. 

\item $\sqrt{|J|} \leq m$ and $ 8\pi |J| \leq A $: the angular momentum appears
  as lower bound and the area appears as upper bound.

\end{enumerate}
The mathematical methods used to study these two groups are, up to now, very
different. This review is mainly concerned with the second group. 

Finally, we mention that for the Kerr black hole there exists a remarkable
equality of the form $(8\pi J)^2 =A^+A^-$, where $A^+$ and $A^-$ denote the
areas of event and Cauchy horizon (see figure \ref{fig:kerr-diag}). This
equality has been proved for general stationary spacetimes in the following
series of articles \cite{Ansorg:2009yi} \cite{Hennig:2009aa}
\cite{Ansorg:2008bv}. It has recently received considerable attention in the
string community (see \cite{Cvetic:2010mn} and \cite{Visser:2012wu} and
references therein). The key property used in these studies is that the product
of horizon areas is independent of the mass of the black hole. It is
interesting to note that there exists, up to now, no generalization of this
kind of equality (or a related inequality) to the dynamical regime.

\section{Theorems}
\label{sec:theorems}
The Penrose inequality
\begin{equation}
  \label{eq:9}
  \sqrt{\frac{A}{16\pi}} \leq m \quad (=\text{Schwarzschild}),
\end{equation}
has been intensively studied. It is a very relevant geometric inequality for
black holes since it is valid without any symmetry assumption.  For a
comprehensive review on this subject see \cite{Mars:2009cj}. The most important
results concerning this inequality are the proofs of Huisken-Ilmanen
\cite{Huisken01} and Bray \cite{Bray01} for the Riemannian case. The general
case remains open. Also, there is up to now no result concerning the Penrose
inequality with angular momentum (\ref{eq:pendj}) discussed in the previous
section.

In the following we present a sample of the main results concerning
inequalities (\ref{eq:mjdd}) and (\ref{eq:JAd}) that have been recently proved.

For the global inequality (\ref{eq:mjdd}) we have the following theorem. 
\begin{theorem}
\label{t:1}
  Consider an axially symmetric, vacuum, asymptotically flat
  and  maximal initial data set with two asymptotics ends.
  Let $m$ and $J$ denote the total mass and angular momentum at one of the
  ends. Then, the following inequality holds
\begin{equation}
\label{eq:60}
\sqrt{|J|} \leq  m \quad (= \text{Extreme Kerr}).
\end{equation}
\end{theorem}
For the precise definitions, fall off conditions an assumptions on the
initial data we refer to original articles cited bellow. 

The first proof of the global inequality (\ref{eq:60}) was provided in a series
of articles \cite{Dain05c}, \cite{Dain05d}, \cite{Dain05e} which end up in the
global proof given in \cite{Dain06c}.  The proof is based on a variational
characterization of the extreme Kerr initial data.  In \cite{Chrusciel:2007dd}
and \cite{Chrusciel:2007ak} the result was generalized and the proof
simplified. In \cite{Chrusciel:2009ki} \cite{Costa:2009hn} the charge was
included. In \cite{Schoen:2012nh} relevant improvements on the rigidity
statements were made. In particular in that article it was proved the first
rigidity result including charge and a measure of the distance to extreme Kerr
black hole was introduced. In \cite{zhou12} the result was proved with the
maximal condition replaced by a small trace assumption for the second
fundamental form of the initial data. Related results concerning the
force between black holes were proved in \cite{Clement:2012np}.
Finally, the mass formula and the variational techniques involved in the proof of the inequality (\ref{eq:60})  were very recently used to study the linear stability of the extreme Kerr black hole   \cite{Dain:2014iba}.

\begin{figure}
\begin{center}
\includegraphics[width=6cm]{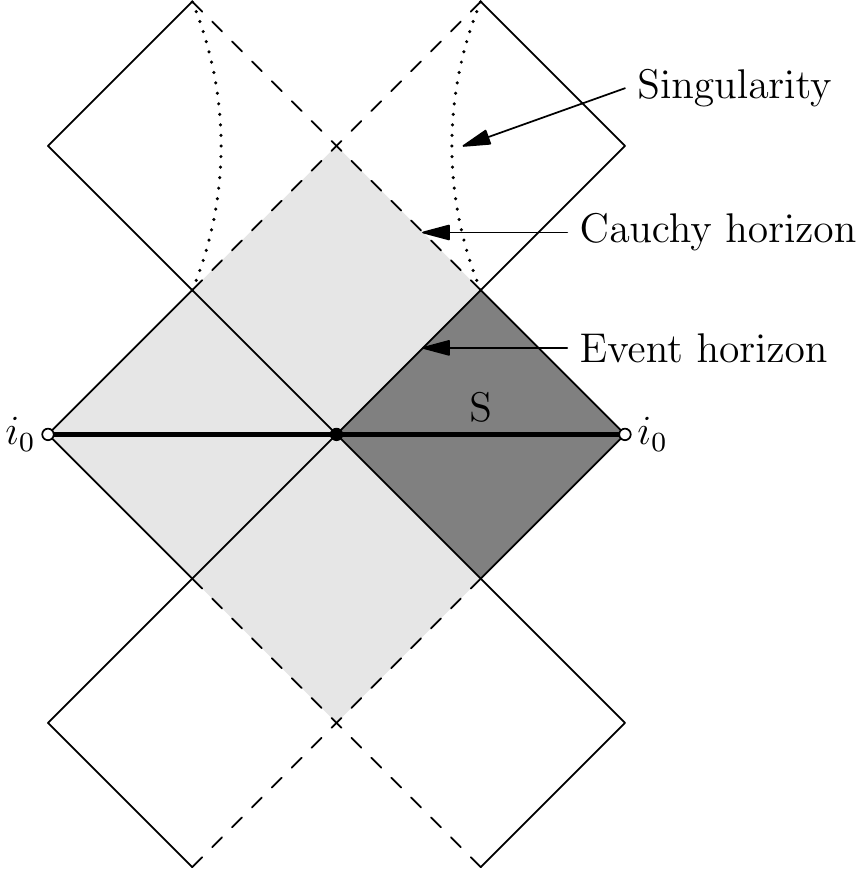} 
\end{center}
\caption{Conformal diagram of the non-extreme Kerr black hole. The points $i_0$
represent spacelike infinity. The surface $S$ have two identical asymptotically flat
ends $i_0$.}
\label{fig:kerr-diag}
\end{figure}

\begin{figure}
  \centering
\includegraphics[width=5cm]{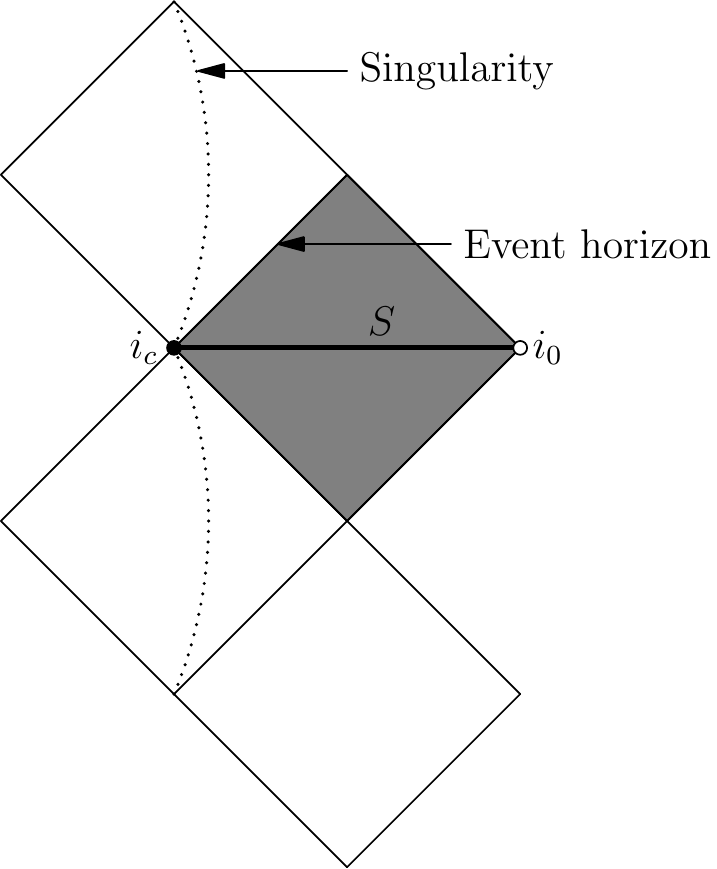} 
\caption{Conformal diagram of the extreme Kerr black hole. The point $i_0$
  represents spacelike infinity, the point $i_c$ represent the cylindrical
  end. The surface $S$ has one asymptotically flat end $i_0$ and one
  cylindrical end $i_c$.}
\label{fig:kerr-extrem-diag}
\end{figure}

\begin{figure}
  \centering
  \includegraphics[width=0.3\textwidth]{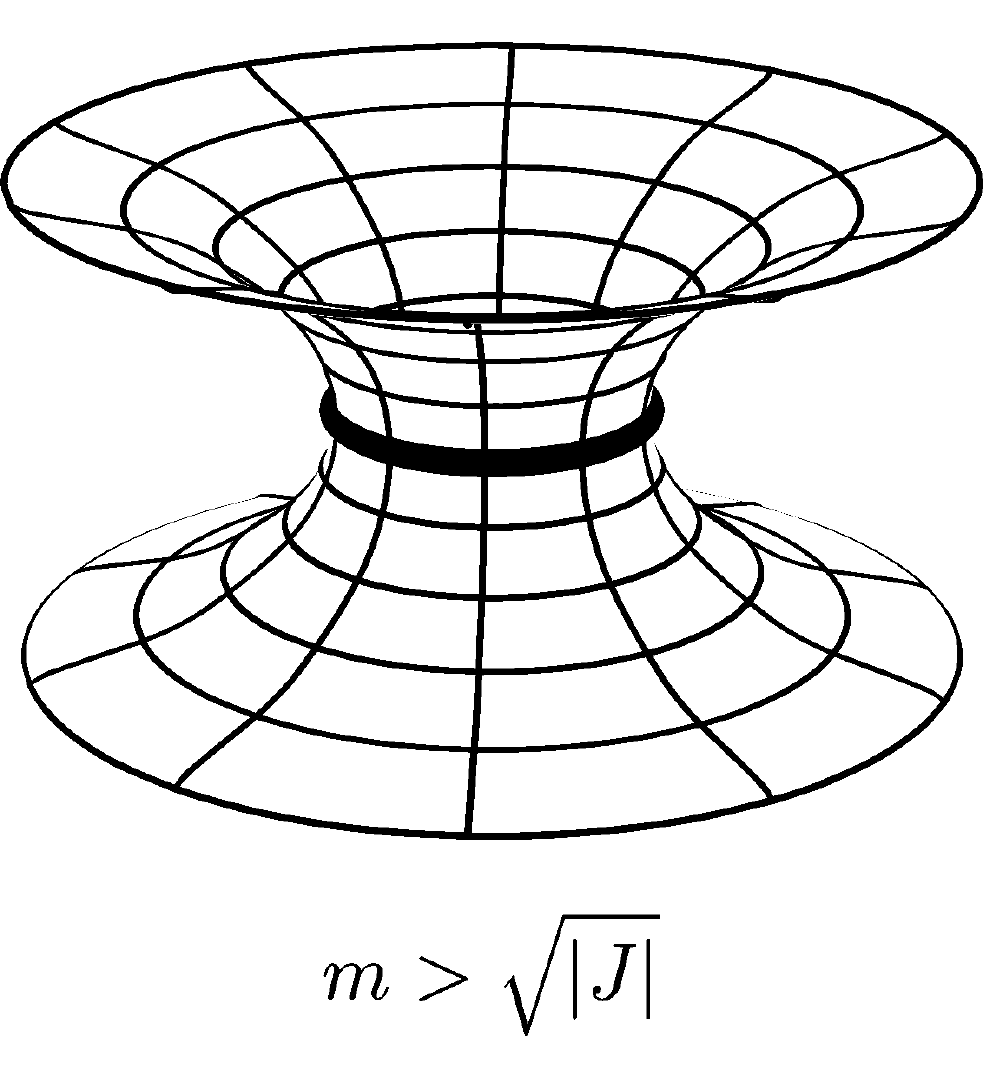}
\hspace{2cm}
 \includegraphics[width=0.35\textwidth]{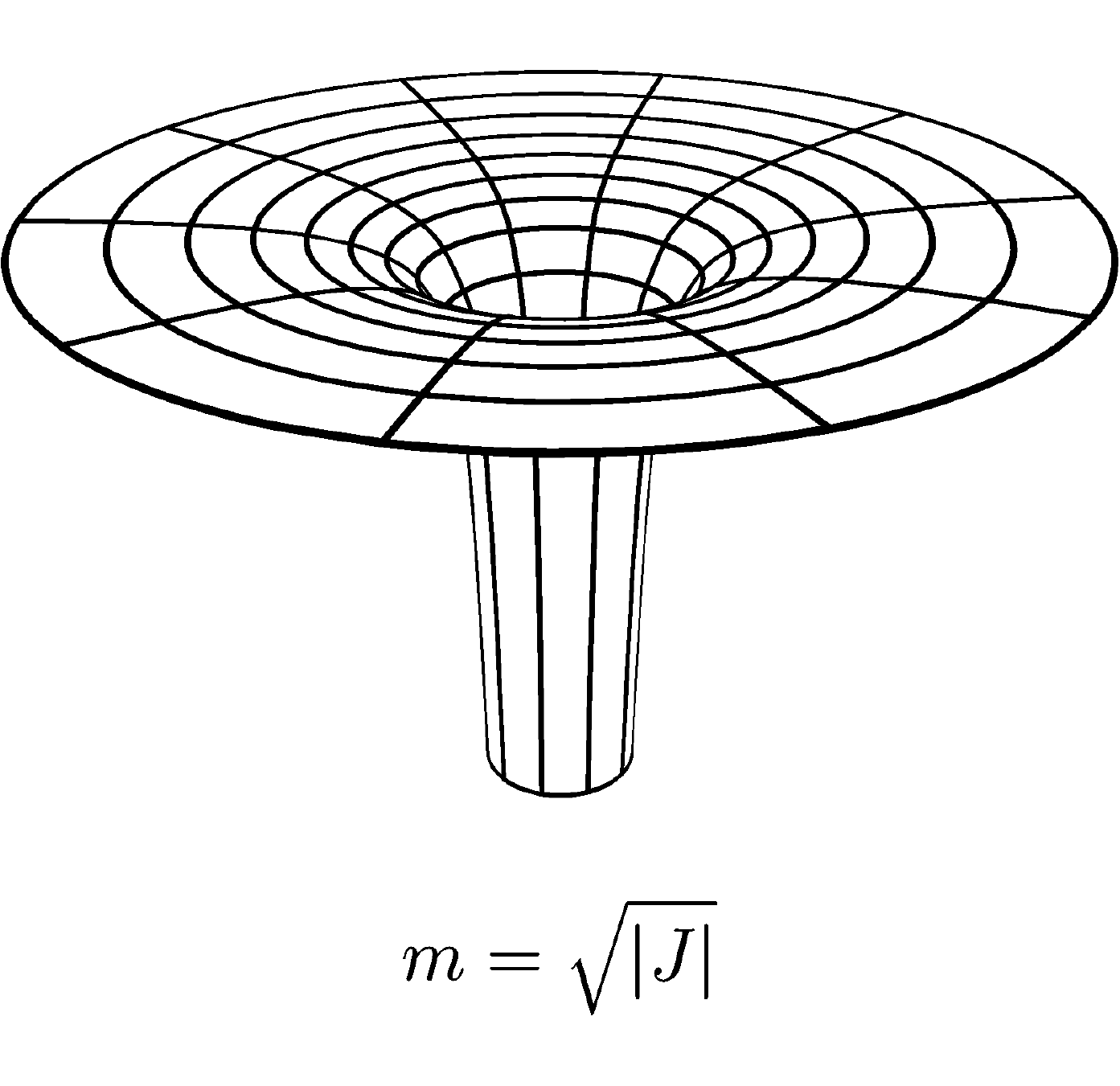} 
 \caption{On the left, an the initial data with two asymptotically flat ends,
   like the non-extreme Kerr black holes. For these data the strict inequality
   holds. On the right, the data of extreme Kerr black hole, with one
   asymptotically flat and one cylindrical end. For this data the equality
   holds.}
  \label{fig:non-extreme-id}
\end{figure}

Under the hypothesis of theorem \ref{t:1} (namely, vacuum and axial symmetry)
the angular momentum is defined as conserved quasi-local integral. In
particular, if the topology of the manifold is trivial (i.e. $\Rt$), then the
angular momentum is zero and hence theorem \ref{t:1} reduces to the positive
mass theorem. In order to have non-zero angular momentum we need to allow
non-trivial topologies, for example manifolds with two asymptotic ends as it is
the case in theorem \ref{t:1}.  An important initial data set that satisfies
the hypothesis of the theorem is provided by an slice $t=constant$ in the Kerr
black hole in the standard Boyer-Lindquist coordinates, see figures
\ref{fig:kerr-diag} and \ref{fig:kerr-extrem-diag}. The non-extreme initial
data have a different geometry as the extreme initial data. The former are
asymptotically flat at both ends. In contrast, extreme initial data, which
reach the equality in (\ref{eq:60}), have one asymptotically flat end and one
cylindrical end, see figure \ref{fig:non-extreme-id}. That geometry represents
the ``optimal shape'' with respect to the inequality (\ref{eq:60}). Figure
\ref{fig:non-extreme-id} is the analog of figure \ref{fig:1} for the
geometrical inequality (\ref{eq:60}).

Regarding the quasi-local inequality (\ref{eq:JAd}) we have the following result. 
\begin{theorem}
\label{t:2}
  Given an axisymmetric closed marginally trapped and stable  surface $\Su$, in a
  spacetime with non-negative cosmological constant and fulfilling the dominant
  energy condition, it holds the inequality
\begin{equation}
\label{eq:JAt}
8\pi |J| \leq  A \quad (= \text{Extreme Kerr throat}),
\end{equation}
where $A$ and $J$  are the area and angular momentum  of $\Su$. 
\end{theorem}
This is a pure spacetime and local result. That is, there is no mention of a
three-dimensional initial hypersurface where the two-dimension surface $\Su$ is
embedded. Axisymmetry is only imposed on $\Su$. Moreover, this theorem does not
assume vacuum.  The matter fields can have also angular momentum and it can be
transferred to the black hole, however the inequality (\ref{eq:JAt}) remains
true even for that case.  It is important to note that the angular momentum that
appears in (\ref{eq:JAt}) is the gravitational one (i.e. the Komar integral).
In fact this inequality is non-trivial even for the Kerr-Newman black hole, see
the discussion in \cite{dain12}.

Theorem \ref{t:2} has the following history. The quasi-local inequality
(\ref{eq:JAt}) was first conjectured to hold in stationary spacetimes
surrounded by matter in \cite{Ansorg:2007fh}. In that article the extreme limit
of this inequality was analyzed and also numerical evidences for the validity
in the stationary case was presented (using the numerical method and code
developed in \cite{Ansorg05}). In a series of articles \cite{hennig08}
\cite{Hennig:2008zy} the inequality (\ref{eq:JAt}) (including also the
electromagnetic charge) was proved for that class of stationary black holes.
See also the review article \cite{Ansorg:2010ru}.

In the dynamical regime, the inequality (\ref{eq:JAt}) was conjectured to hold
in \cite{dain10d} based on the heuristic argument mentioned in section
\ref{sec:physical-picture}. In that article also the main relevant techniques
for its proof were introduced, namely the mass functional on the surface and
its connections with the area.  A proof (but with technical restrictions) was
obtained in \cite{Acena:2010ws} \cite{Clement:2011kz}. The first general and
pure quasi-local result was proven in \cite{Dain:2011pi}, where the relevant
role of the stability condition for minimal surfaces was pointed out. The
generalization to trapped surfaces and non-vacuum has been proved in
\cite{Jaramillo:2011pg}. The electromagnetic charge was included in
\cite{Clement:2011np} and \cite{Clement:2012vb}. This inequality has been extended to
higher dimensions in \cite{Hollands:2011sy} and \cite{Paetz:2013rka}. In
\cite{Yazadjiev:2012bx} \cite{Yazadjiev:2013hk} and \cite{Fajman:2013ffa} it
has been also extended to Einstein-Maxwell dilaton gravity.  In
\cite{Reiris:2013jaa} related inequalities that involve the shape of the black
hole were proved.

\begin{figure}
  \centering
  \includegraphics[width=0.2\textwidth]{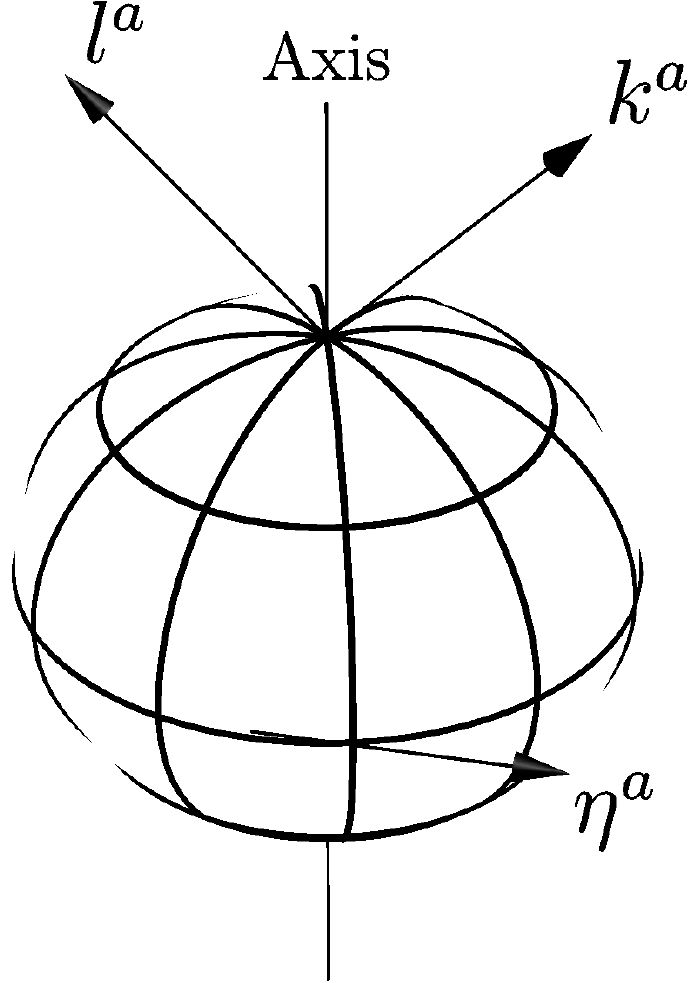} 
  \caption{Axially symmetric two-surface. The axial Killing vector $\eta$ is
    tangent to the surface. The null vectors  $\ell^a$ and $k^a$  are normal to ${\cal S}$}
  \label{fig:axial-2s}
\end{figure}

To describe the concept of stable trapped surface (this condition was first
introduced in \cite{andersson08}) used in theorem \ref{t:2} let us consider an
axially symmetric closed two-surface $\Su$ with the topology of a two-sphere.
The surface $\Su$ is embedded in the spacetime.  Let $\ell^a$ and $k^a$ be null
vectors spanning the normal plane to $\Su$ and normalized as $\ell^a k_a =
-1$, see figure \ref{fig:axial-2s}. The expansion is defined by
$\theta^{(\ell)}= \nabla_a\ell^a$, where $\nabla$ is the spacetime
connection. The surface $\Su$ is marginally trapped if $\theta^{(\ell)}=0$.
Given a closed marginally trapped surface $\Su$ we will refer to it as
spacetime stably outermost if there exists an outgoing ($-k^a$-oriented) vector
$X^a= \gamma \ell^a - \psi k^a$, with $\gamma\geq0$ and $\psi>0$, such that the
variation of $\theta^{(\ell)}$ with respect to $X^a$ fulfills the condition
\begin{equation}
\label{e:stability_condition}
\delta_X \theta^{(\ell)} \geq 0.
\end{equation}
Here $\delta$ denotes a variation operator associated with a deformation of the
surface $\Su$ (c.f. for example \cite{Booth:2006bn}
\cite{andersson08})). For maximal initial data the stability condition
(\ref{e:stability_condition}) is closed related with the stability condition
for minimal surfaces (see \cite{Dain:2011kb}, \cite{Jaramillo:2011pg}). The
stability of a minimal surface is the requirement that the area is a local
minimum. 

\begin{figure}
  \centering
  \includegraphics[width=0.5\textwidth]{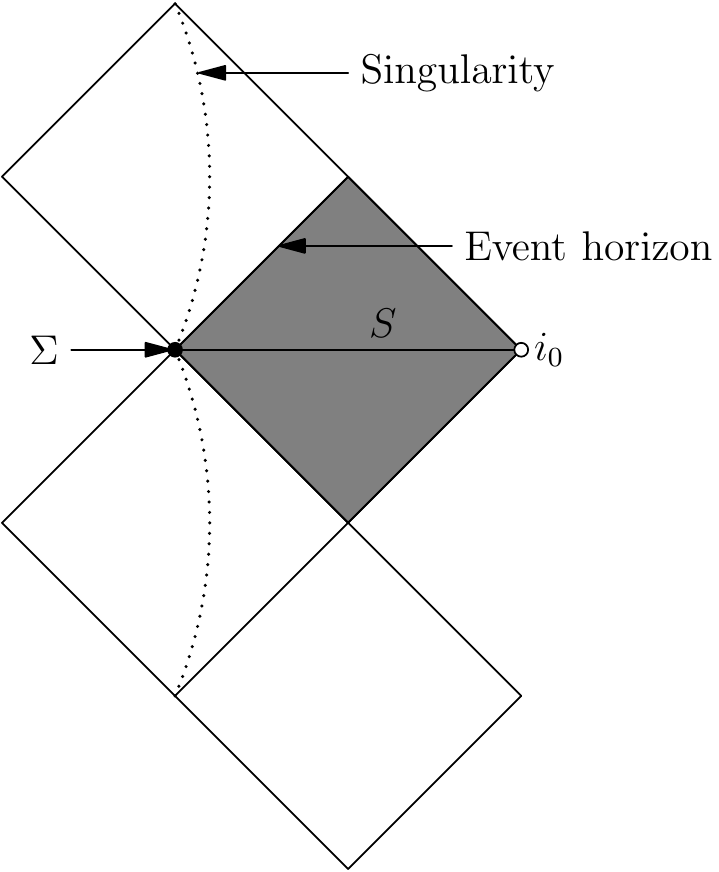}
  \caption{Location of the extreme Kerr throat surface $\Su$ in the spacetime. }
  \label{fig:throat-cd}
\end{figure}

\begin{figure}
  \centering
  \includegraphics[width=0.4\textwidth]{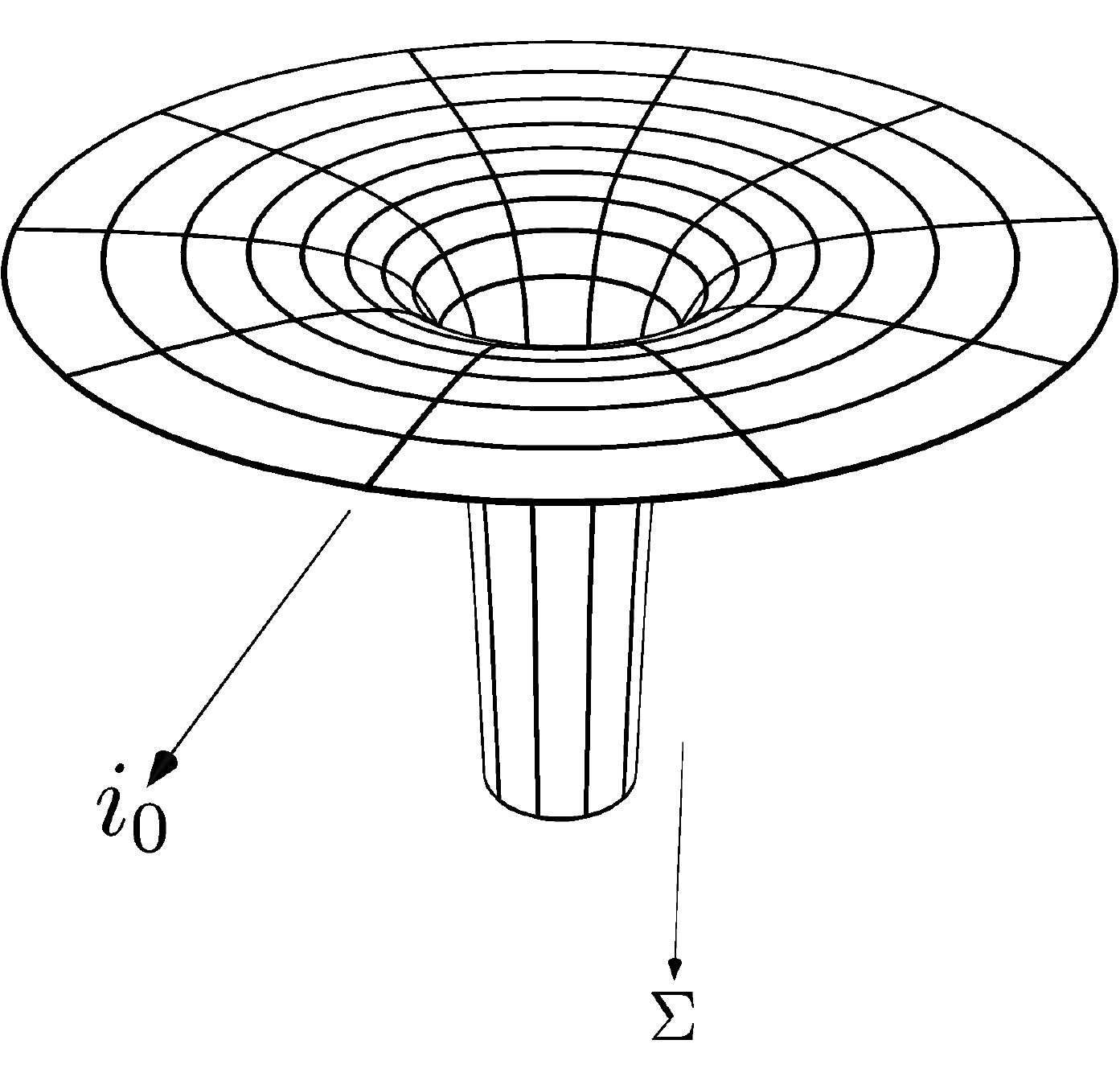}
  \caption{Location of the extreme Kerr throat surface $\Su$ on the initial data. }
  \label{fig:throat-id}
\end{figure}

\begin{figure}
  \centering
  \includegraphics[width=0.3\textwidth]{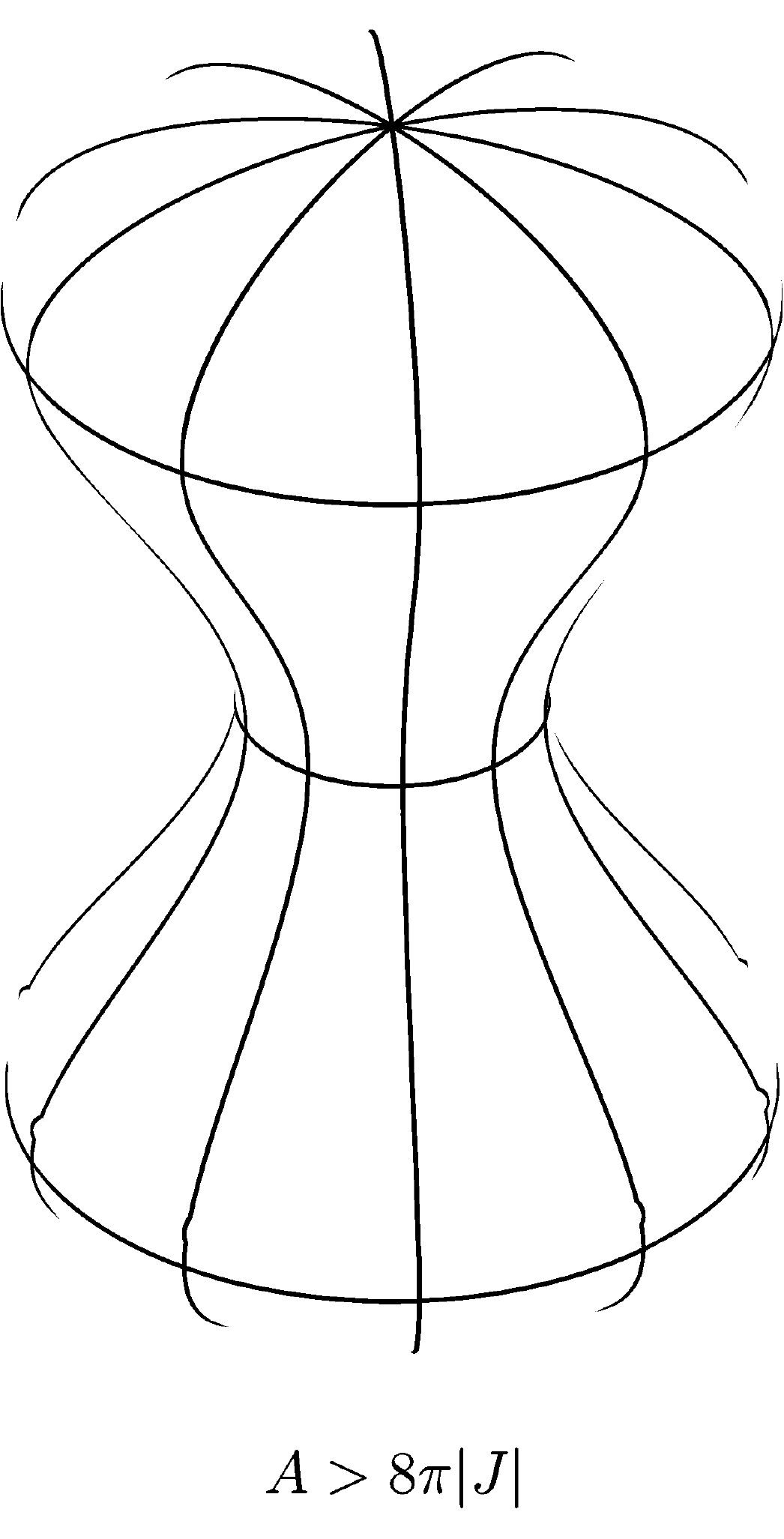}
\hspace{2cm}
\includegraphics[width=0.3\textwidth]{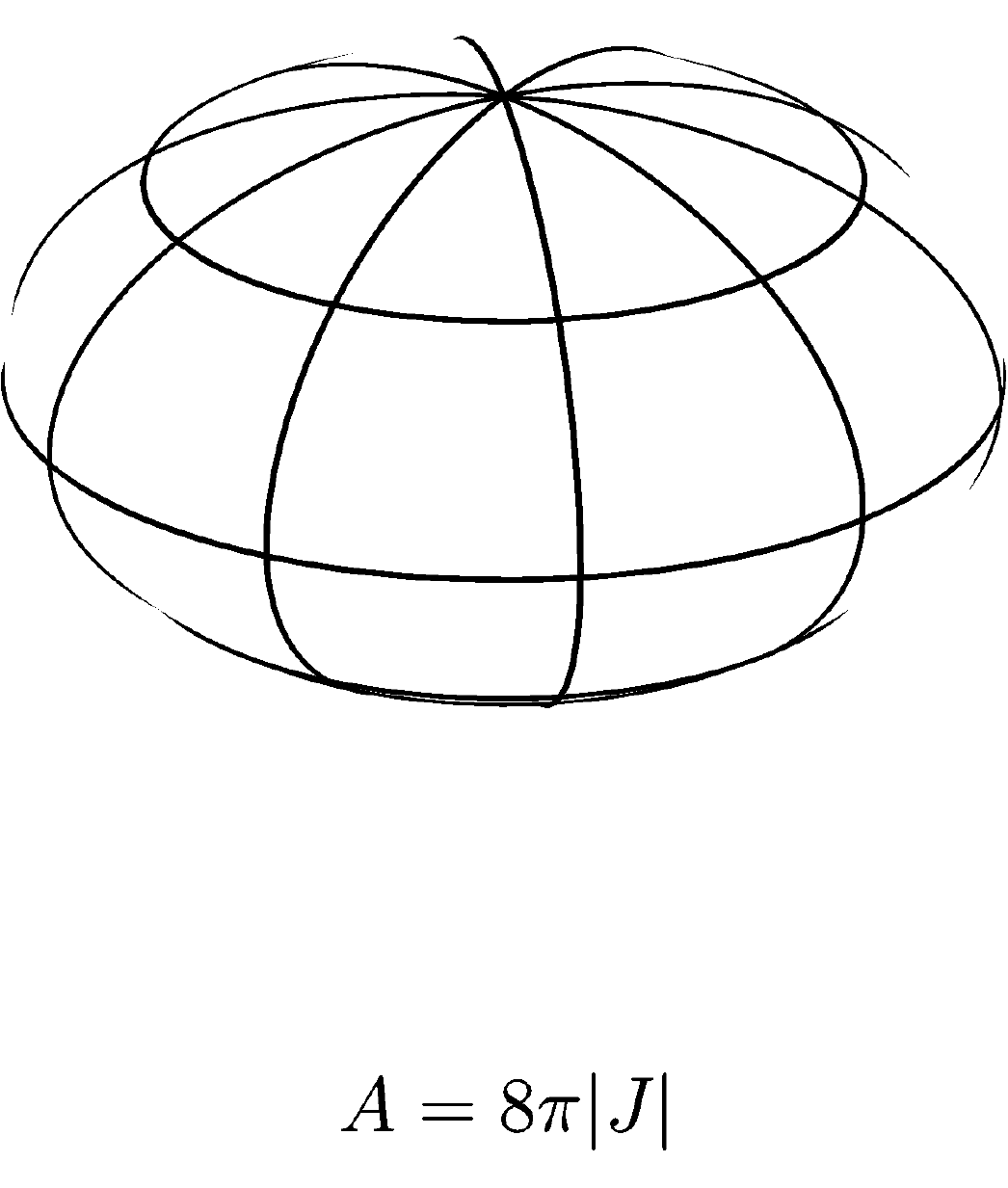}   
\caption{On the left, an arbitrary axially symmetric stable two surface. For
  this kind of surface the strict inequality holds. On the right,  the extreme
  throat sphere, where the equality holds.}
  \label{fig:arb-st}
\end{figure}

The extreme throat geometry, with angular momentum $J$, was defined in
\cite{dain10d} (see also \cite{Acena:2010ws} and \cite{Dain:2011pi}). This
concept captures the local geometry near the horizon of an extreme Kerr black
hole.  The extreme throat is the asymptotic limit in the cylindrical end of an
extreme Kerr black hole, see figure \ref{fig:throat-cd} and
\ref{fig:throat-id}. Both the intrinsic and extrinsic geometry of this surface
are fixed. For example, it has an intrinsic metric given by
\begin{equation}
\label{eq:gamma0}
|J| \left( (1+\cos^2\theta) d\theta^2+ \frac{4\sin^2\theta}{(1+\cos^2\theta)} d\phi^2 \right).
\end{equation}
It is an oblate sphere with respect to the axis of rotation (see figure
\ref{fig:arb-st}, on the right).

The extreme Kerr throat achieve the equality in (\ref{eq:JAt}), this surface
has the ``optimal shape'' with respect this inequality. It has also a
variational characterization. Figure \ref{fig:arb-st} is the analog of figures
\ref{fig:1} and  \ref{fig:non-extreme-id} for inequality   (\ref{eq:JAt}). 

The results in theorem \ref{t:2} has been used  in a  recent non-existence proof of stationary black holes binaries \cite{Neugebauer:2013ee} \cite{Neugebauer:2011qb} \cite{Chrusciel:2011iv}. 

The rigidity statement in theorem \ref{t:2} (namely that the equality in
\eqref{eq:JAt} implies that the surface is an extreme Kerr throat) has been
proved in a different context: for extreme isolated horizon and near-horizon
geometries of extremal black holes in \cite{Hajicek:1974oua},
\cite{Lewandowski:2002ua} and \cite{Kunduri:2008rs}, see also the review
article \cite{lrr-2013-8} and reference therein.

\section{Open problems and recent results on bodies} 
\label{sec:open-problems-recent}
In this final section I would like to present the main open problems regarding
the black holes geometrical inequalities discussed in the previous sections.
My aim is to present open problems which are relevant (and probably involve the
discovery of new techniques) and at the same time they appear feasible to
solve. For more details see the review article \cite{dain12}. The open problem
mentioned there regarding the inclusion of the electric charge in the
quasi-local inequality (\ref{eq:JAt}) have been solved
\cite{Clement:2011np} \cite{Clement:2012vb}.

For the global inequality (\ref{eq:60}) 
there are two main open problems, which involve generalizations of the  assumptions in
theorem \ref{t:1}:
\begin{itemize}
\item Remove the maximal condition.

\item Generalization for asymptotic flat manifolds with multiple ends. 
\end{itemize}

Concerning the maximal condition, as we mention above, in a recent article
\cite{zhou12} this assumption have been replaced by a small trace
condition. See also the discussion in \cite{dain12}.  The most relevant open
problem is the second one. The physical heuristic argument presented in section
\ref{sec:physical-picture} applies to that case and hence there little doubt
that the inequality holds.  This problem is related with the uniqueness of the
Kerr black hole with degenerate and disconnected horizons. It is probably a
hard problem. There are very interesting partial results in
\cite{Chrusciel:2007ak} and also numerical evidences in \cite{Dain:2009qb}.

Probably the most important open problem for geometrical inequalities for
axially symmetric black holes is the following:
\begin{itemize}
\item Prove the Penrose inequality with angular momentum (\ref{eq:6}). 
\end{itemize}
We mention in section \ref{sec:physical-picture} that there is a clear physical
connection between the global inequality (\ref{eq:60}) and the Penrose
inequality with angular momentum in axial symmetry (\ref{eq:6}). However, the
techniques used to prove the inequality (\ref{eq:60}) are very different than
the one used to prove the classical Penrose inequality (\ref{eq:9}) (see the
discussion in \cite{dain12}).

For the quasi-local inequality (\ref{eq:JAt}) the two main problems are the
following:

\begin{itemize}
\item A generalization of the inequality  (\ref{eq:JAt})   without axial  symmetry.

\item A generalization of the inequality   (\ref{eq:JAt})  for ordinary bodies. 

\end{itemize}
The problem of finding versions of inequality (\ref{eq:JAt}) without any
symmetry assumption, in contrast with the other open problems presented above,
is not a well-defined mathematical problem since there is no unique notion of
quasi-local angular momentum in the general case. However, exploring the scope
of the inequality in regions close to axial symmetry (in some appropriate
sense) can perhaps provide such a notion. From the physical point of view, we
do not see any reason why this inequality should only hold in axial
symmetry. Note that the global inequality (\ref{eq:60}) only holds in axial
symmetry. This is clear from the physical point of view (see the discussion in
\cite{dain12}) and in \cite{huang11} highly non-trivial counter examples have
been constructed.

\begin{figure}
  \centering
  \includegraphics[width=2.6cm]{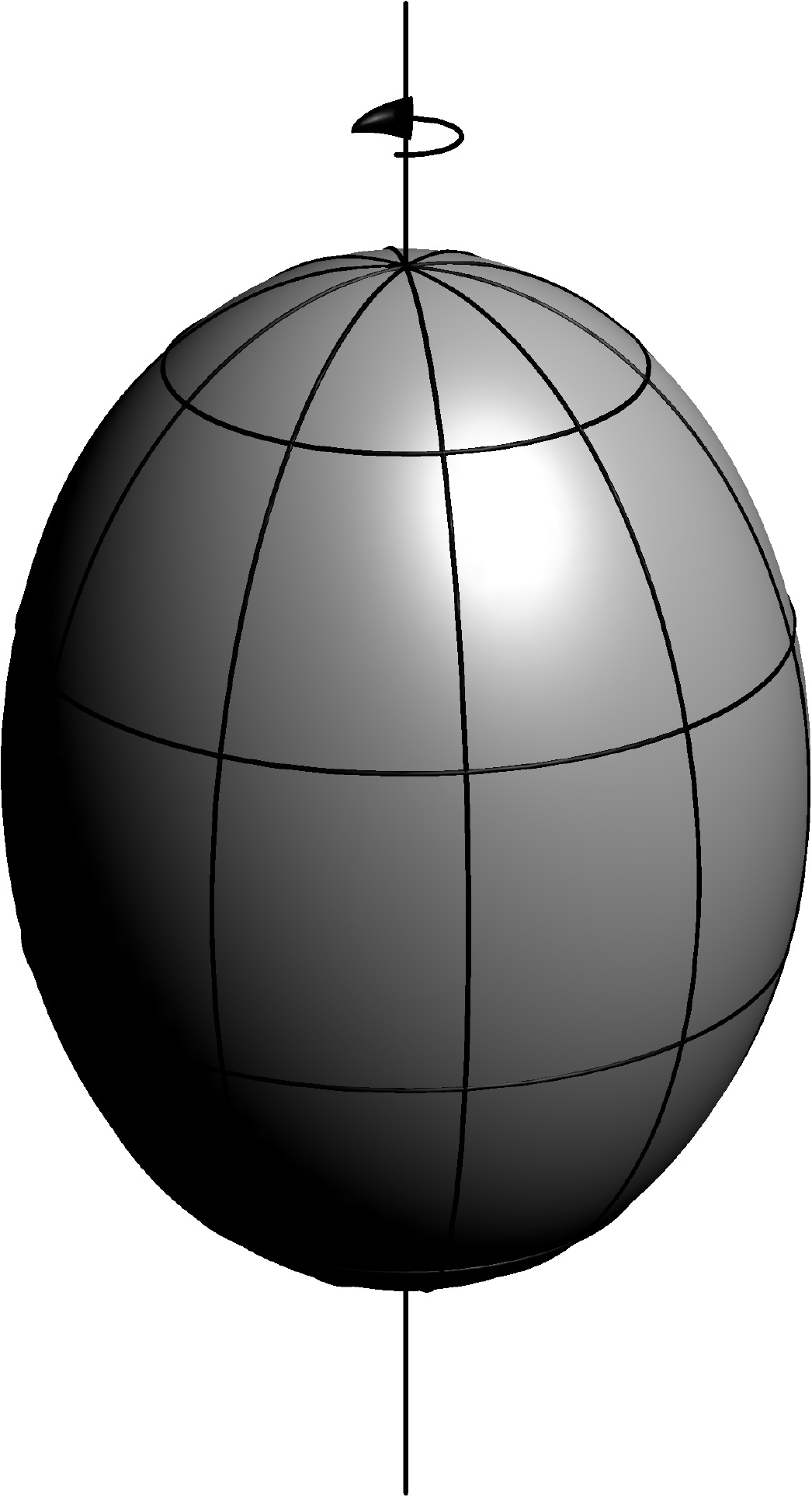} 
  \caption{Axially symmetric rotating body.}
  \label{fig:body}
\end{figure}

Finally, concerning the second problem there have been recently some results in
\cite{Dain:2013gma}. Consider a rotating body $\dom$ with angular momentum
$J(\dom)$, see figure \ref{fig:body}.  Let $\Si(\dom)$ be a measure (with units of length) of the size
of the body.

In \cite{Dain:2013gma},  the following universal
inequality for all bodies is conjectured
\begin{equation}
  \label{eq:22}
  \Si^2(\dom) \apprge  \frac{G}{c^3} |J(\dom)|, 
\end{equation} 
where $G$ is the gravitational constant and $c$ the speed of light. The symbol
$\apprge$ is intended as an order of magnitude, the precise universal
(i.e. independent of the body) constant will depend on the definition of
$\Si$. We have reintroduced in (\ref{eq:22}) the fundamental constants in order
to make more transparent the discussion bellow.

The arguments in support of the inequality (\ref{eq:22}) are based in the
following three physical principles:
 \begin{itemize}
\item[(i)] The speed of light $c$ is the maximum speed.

\item[(ii)] For bodies which are not contained in a black hole the following
  inequality holds
  \begin{equation}
    \label{eq:2b}
    \Si(\dom) \apprge\frac{G}{c^2}  m(\dom),
  \end{equation}
  where $m(\dom)$ is the mass of the body. 

\item[(iii)] The inequality (\ref{eq:22})   holds for black holes.
\end{itemize}
Let us discuss these assumptions. Item (i) is clear. Item (ii) is called the
\emph{trapped surface conjecture} \cite{Seifert79}.  Essentially, it says that
if the reverse inequality as in (\ref{eq:2b}) holds then a trapped surface
should enclose $\dom$. That is: if matter is enclosed in a sufficiently small
region, then the system should collapse to a black hole.  This is related with
the \emph{hoop conjecture} \cite{thorne72} (see also \cite{Wald99}
\cite{PhysRevD.44.2409} \cite{Malec:1992ap} ).  The trapped surface conjecture
has been proved in spherical symmetry \cite{Bizon:1989xm} \cite{Bizon:1988vv}
\cite{Khuri:2009dt} and also for a relevant class of non-spherical initial data
\cite{Malec:1991nf}. The general case remains open but it is expected that some
version of this conjecture should hold.

Concerning item (iii), the area $A$ is a measure of the size of a trapped
surface, hence the inequality (\ref{eq:JAt}) represents a version of
\eqref{eq:22} for axially symmetric black holes. If we include the physical
constants, this inequality has the form
\begin{equation}
  \label{eq:5}
  A\geq8\pi\frac{G}{c^3} |J|.
\end{equation}
In fact the inequality (\ref{eq:5}) was the inspiration for the inequality
 (\ref{eq:22}).  A possible generalization of (\ref{eq:5}) for bodies is to take
 the area $A(\partial \dom)$ of the boundary $\partial \dom$ of the body $\dom$
 as measure of size. But unfortunately the area of the boundary is not a good
 measure of the size of a body in the presence of curvature. In particular, an
 inequality of the form $A(\partial \dom) \apprge G c^{-3} |J(\dom)| $ does not
 holds for bodies. The counter example is essentially given by a rotating torus
 in the weak field limit, with large major radius and small minor radius.  The
 details of this calculation will be presented in \cite{Anglada13}.

 Using the three physical principles (i), (ii) and (iii) in \cite{Dain:2013gma}
 it is argued that the inequality (\ref{eq:22}) should hold. One of the main
 difficulties in the study of inequalities of the form \eqref{eq:22} is the very
 definition of the measure of size. In fact, despite the intensive research on
 the subject, there is no know universal measure of size such that the trapped
 surface conjecture (or, more general, the hoop conjecture) holds (see the
 interesting discussions in \cite{Malec:1992ap} \cite{Gibbons:2012ac}
 \cite{Senovilla:2007dw} \cite{Reiris:2013jaa}). However, the remarkable point
 is that in order to find an appropriate measure of size $\Si$ such that
 \eqref{eq:22} holds it is not necessary to prove first \eqref{eq:2}, and hence
 we do not need to find the relevant measure of mass $m(\dom)$ for the trapped
 surface conjecture. In \cite{Dain:2013gma} a size measure is proposed and for
 that measure the following version of the inequality \eqref{eq:22} has been
 proved for constant density bodies. This theorem is a consequence of the
 Schoen-Yau theorem \cite{schoen83d}.

\begin{figure}
   \centering
   \includegraphics[width=3.2cm]{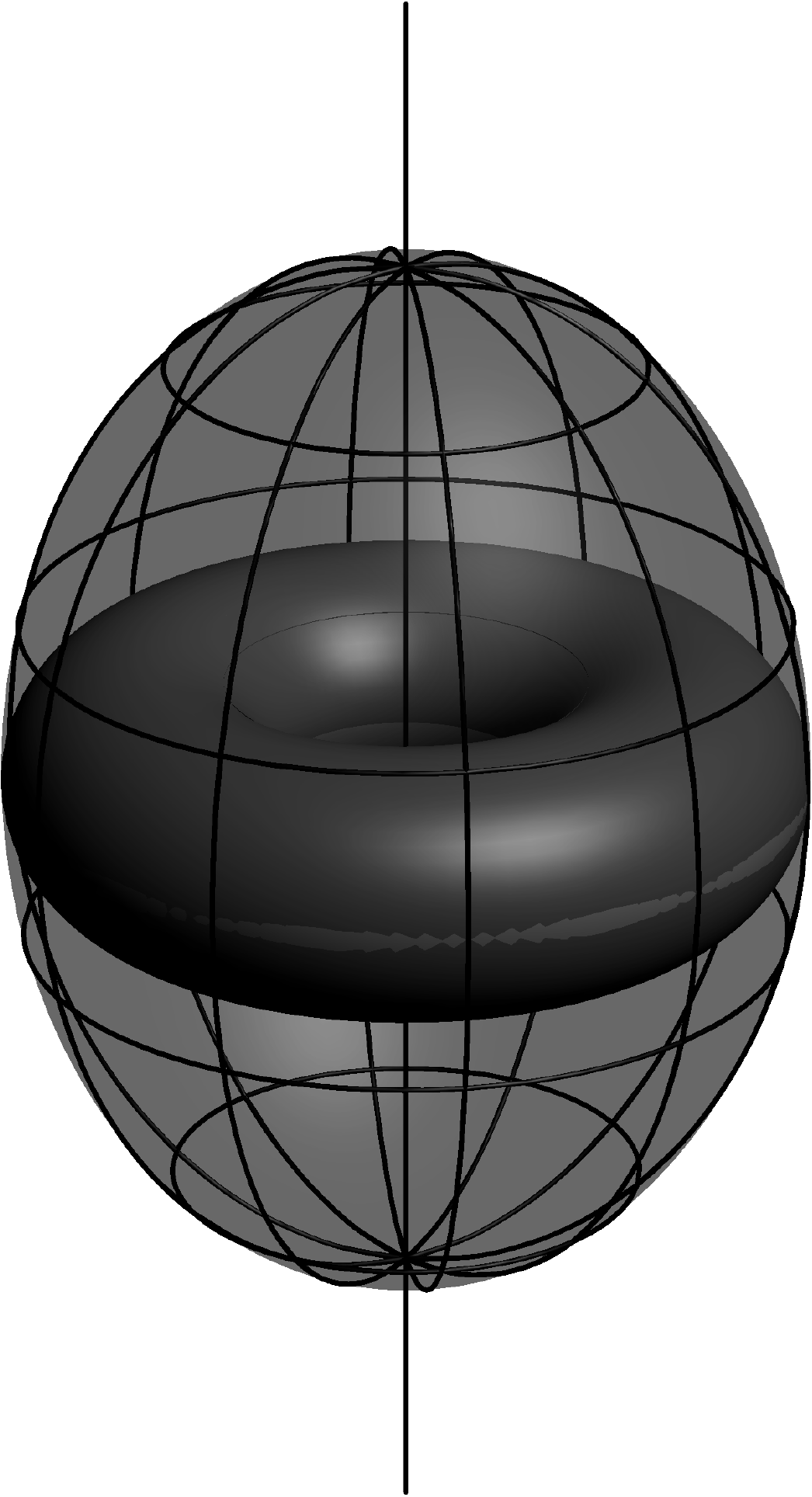}\hspace{2cm}
    \includegraphics[width=1.8cm]{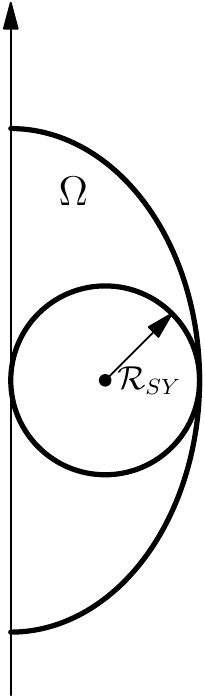} 
    \caption{On the left, the Schoen-Yau $\ra$ radius for a body is defined in
      terms of the biggest embedded torus. On the right, the same torus is
      showed on the plane orthogonal to the axial Killing vector.  On that plane
      the torus is a circle, and the radius $\ra$ is related to the radius of
      the biggest embedded circle.}
   \label{fig:sy-torus}
 \end{figure}

 \begin{theorem}
 \label{t:3}
 Consider a maximal, axially symmetric, initial data set that satisfy the
 dominant energy condition.  Let $\dom$ be an open set on the data. Assume that
 the energy density is constant on $\dom$.  Then the following inequality holds
  \begin{equation}
    \label{eq:7d}
  \Si^2(\dom) \geq  \frac{24}{\pi^3}\frac{G}{c^3} |J(\dom)|.
  \end{equation}
 \end{theorem}
 The definition of the radius $\Si$ in (\ref{eq:7d}) is as follow.  Let
 $\ra(\dom)$ be the Schoen-Yau radius defined in \cite{schoen83d}. This radius
 is expressed in terms of the largest torus that can be embedded in $\dom$. See
 figure \ref{fig:sy-torus}.

 Consider a region $\dom$ with a Killing vector
 $\eta^i$ with norm $\lambda$,  we define the radius $\Si$ by
 \begin{equation}
   \label{eq:8}
   \Si(\dom) =  \frac{2}{\pi} \frac{\left(\int_\dom \lambda  \right)^{1/2}}{\ra(\dom)}. 
 \end{equation}
The definition of the radius (\ref{eq:8}) is, no doubt, very involved. It is not expected to be the optimal size measure for a body.  It should be considered, together with theorem \ref{t:3},  as an example where the conjecture (\ref{eq:22}) can be proved  with the current available mathematical techniques.
For examples and further discussion on this radius  we refer to \cite{Dain:2013gma}.

 \vspace{1cm}

 This article is based on the longer review article \cite{dain12}, we refer to
 that article for more details. The two main differences with respect to
 \cite{dain12} are the following.  First, several new results appeared after the
 publication of \cite{dain12}.  These results have been included here. Second,
 the physical arguments in section \ref{sec:physical-picture} have been
 significantly improved and clarified, based on the discussion in
 \cite{Dain:2013qia}.

 \section*{Acknowledgements}
   This work was supported  by grant PICT-2010-1387 of CONICET (Argentina) and
   grant Secyt-UNC (Argentina).
  

\end{document}